\documentclass[a4paper]{jpconf}
\usepackage{graphicx}

\begin{document}
\title{Re II and Other Exotic Spectra in HD 65949}

\author{C R Cowley$^1$, S Hubrig$^2$, G M Wahlgren$^3$}

\address{$^1$Dept. of Astronomy, Univ. of Michigan,
Ann Arbor, MI 48109-1042, USA}
\address{$^2$European Southern Observatory, Casilla 19001,
Santiago 19, Chile}
\address{$^3$Catholic Univ. of America,
620 Michigan Ave NE, Washington, DC 20064, USA}

\ead{$^1$cowley@umich.edu, $^2$shubrig@eso.org,
$^3$Glenn.M.Wahlgren@nasa.gov}

\newcommand{\ion}[2]{\textup{#1}\,\textsc{#2}}

\begin{abstract}
Powerful astronomical spectra reveal an urgent need for additional
work on atomic lines, levels, and oscillator strengths.  The star
HD 65949 provides some excellent examples of species rarely identified in
stellar spectra.   For example, the \ion{Re}{ii}
spectrum is well developed, with 17 lines between
3731 and 4904 \AA, attributed wholly or partially to \ion{Re}{ii}.
Classifications and oscillator strengths are lacking for
a number of these lines.  The spectrum of \ion{Os}{ii} is well
identified.  Of 14 lines attributed wholly or partially
to \ion{Os}{ii}, only one has an entry in the VALD database.  We
find strong evidence that \ion{Te}{ii} is present.  There are
{\it no} \ion{Te}{ii} lines in the
VALD database.  \ion{Ru}{ii} is clearly present, but oscillator strengths
for lines in the visual are lacking.  There is excellent to
marginal evidence for a number of less commonly identified
species, including \ion{Kr}{ii}, \ion{Nb}{ii},
\ion{Sb}{ii}, \ion{Xe}{ii}, \ion{Pr}{iii}, \ion{Ho}{iii},
\ion{Au}{ii}, and \ion{Pt}{ii} (probably $^{198}$Pt), to be
present in the spectrum of HD 65949.  The line \ion{Hg}{ii} $\lambda$3984
is
of outstanding strength, and all three lines of Multiplet 1
of \ion{Hg}{i} are present, even though the surface temperature
of HD 65949 is relatively high. Finally, we present the case of an
unidentified, 24\,m\AA\, line at 3859.63\,\AA\, which
could be the same feature seen in magnetic CP stars.
It is typically blended with a putative \ion{U}{ii} line
used in cosmochronology.
\end{abstract}

\section{Introduction}
A variety of chemically-peculiar (CP) stars are found on
or near the upper main sequence.  Most of these fall into
one of two main groups: the magnetic silicon stars, or
the non-magnetic mercury-manganese (HgMn) stars.
HD 65949 shares characteristics of both of these groups
(Cowley et al. 2006, henceforth, Paper I).
Intermediate-dispersion spectra of HD 65949 obtained by
Abt \& Morgan (1969) already revealed the outstanding
strength of \ion{Hg}{ii} $\lambda$3984, a signature
feature of the HgMn stars.  An estimate of the mercury
abundance presented in Paper I was higher than any other published
value.  Yet HD 65949 is quite unlike the typical HgMn
stars.  The \ion{Mn}{ii} spectrum was present, but much weaker
than expected for an HgMn star with its surface temperature,
near 13,000K.

Hubrig, et al. (2006) had
measured a  magnetic field, of the order of
few hundred Gauss.  This field is weaker than typical for
a silicon star.  Moreover, two characteristic features
of silicon stars were missing; the spectrum of \ion{Si}{ii} was
not particularly strong, and the second spectra of the
lanthanides were only weakly present.

HD 65949 may be grouped with two other
``misfits'' near its temperature and surface gravity
range: HR 6870 (HD 168733, cf. Muthsam \& Cowley 1984),
and HR 6000 (HD 144667, Castelli \& Hubrig, 2007).
All three spectra are distinct, as can be seen from
Figs. 1 and 3 of Paper I.

In Paper I, several uncommon atomic spectra were found
to be definitely or possibly present in the spectrum of HD 65949.  We
review these in the sections that follow, along with
work done subsequently on the model atmosphere and spectrum
of the star.  We do not claim any ``first'' identifications,
but draw attention to several atomic spectra found in HD 65949
that are rarely found in stellar spectra.

\section{Observations}

This paper is based primarily on one of four spectra
taken with the ESO/FEROS echelle spectrograph
in October 2005.  The
usable wavelength coverage is from 3530 to 9220~\AA,
but the spectra are not of good quality to the violet
of ca. 3590~\AA.  The resolving power is about 48\,000, and
the signal-to-noise (S/N) ratio reaches up to 250.
Wavelength and equivalent width
measurements were
made on the spectrum after it was mildly Fourier-filtered.

\section{An atmospheric model}
\subsection{Basic parameters}

HD 65949 is a mid-range, main-sequence B-type star.  For such
objects, the Balmer line profiles are far more sensitive
to the surface gravity ($\log(g)$) than the effective temperature ($T_{\rm
eff}$).
The effective temperature, obtained from UBV and Str\"{o}mgren
photometry, spans a range from 12\,200 to 13\,600K, depending on
the system and calibration.  Details may be found at
\verb=http://www.astro.lsa.umich.edu/~cowley/hd65949=.
We find good to excellent fits of H$\alpha$, H$\beta$,
and H$\delta$, within this temperature range
for $\log(g) = 4.0$.

In Paper I, we used a $T_{\rm eff}$ of 13\,600K, at the
high end of the photometric values.  We now find that a
lower temperature gives abundance equality
among \ion{Fe}{i,\,ii,\,iii}, and adopt that
value, $T_{\rm eff} = 12\,600$K, along with
with $\log(g) = 4.0$.  In doing this, we assume the iron abundance
is constant throughout the atmosphere.  Much new work indicates
this assumption may not be true, but for the present, it is
a useful, working assumption.

We now obtain the following values for $\log({\rm N_{Fe}/N_{total}})$,
the logarithmic ratio of the iron abundance to the sum of
all abundances: \ion{Fe}{i}, $-$3.95 (21); \ion{Fe}{ii}, $-$4.12 (101);
\ion{Fe}{iii}, $-$3.97 (7). The values in parenthesis indicate
the number of equivalent widths used for each determination.
Oscillator strengths for \ion{Fe}{i,\,ii} were taken from Fuhr \& Wiese
(2006); the VALD values (Kupka, et al. 2000) were used
for \ion{Fe}{iii}.
Our model atmosphere has a helium
abundance of 0.5 solar and an iron abundance of
$\log({\rm N_{Fe}/N_{\rm total}}) = -4.1$,
with the remaining abundances being similar to their solar-system values.

\subsection{The helium and oxygen abundances}
Abundance determinations were not systematically carried out for
all identified spectra.  We mention here two elements whose
abundances can be significant for the structure of the
atmosphere.
We made a rough estimate of the helium abundance
from seven \ion{He}{i} lines, by a line-fitting techniques
using Voigt profiles.
We find log(N$_{\rm He}$/N$_{\rm total}$) = $-$1.40; helium
is about half of its abundance in the sun.
Iron is about a factor of 3 above its solar abundance.
We estimated the
oxygen abundance from 10 \ion{O}{i} lines, excluding the strong
triplet, 7772, 7774, and 7775~\AA, which is sensitive
to NLTE.  Log(N$_{\rm O}$/N$_{\rm total}$) = $-$3.32, which
is very nearly solar (log(N$_{\rm O}$/N$_{\rm total}$)$_\odot$ = 3.38).

\section{The Re II spectrum}

\subsection{Wavelengths and identifications}

Figure 2 of Paper I shows
a feature in the red wing of
the \ion{Hg}{ii} $\lambda$3984 line that was difficult to
identify.  It turned out to be one of the weaker
lines of \ion{Re}{ii} classified by Meggers, Catal\'{a}n,
\& Sales (1958).

\begin{table}[ht]
\caption{\label{tab:re2}Stellar and laboratory Re lines}
\scriptsize
\begin{center}
\begin{tabular}{llll}
\br
$\lambda^* (\AA )$& Cent(i)&\rule{8mm}{0mm}ID$_1$&\rule{8mm}{0mm}ID$_2$ \\
3731.66   & 0.92 &  .66 Re II (10Hcw)     &                      \\
{\bf 3742.30}& 0.66 &  .26 Re II-.133        &  .32 Mo II-.017      \\
3772.99   & 1.00 &  .02 Re II (15cw)      &                      \\
3776.27   & 0.92 &  .30 Re (20c.24Wl)wm,uc&                      \\
3800.93   & 0.87 &  .95 Re II (100cw)     &                      \\
3830.57   & 0.90 &  .52 Re II (8h)        &  .62 Re (10h) wm,uc  \\
3839.54   & 0.99 &  .54 Re II (10h)       &                      \\
3847.69   & 0.93 &  .72 Re II (10Hw)      &                      \\
3858.53   & 0.96 &  .53 Re (5h) wm,uc     &                      \\
3873.50   & 1.00 &  .60 Re (6hw) wm,uc    &                      \\
3898.57   & 0.94 &  .60 Fe II-.247        &  .50 Re (5h) wm,uc   \\
3910.59   & 0.99 &  .59 Re (3h) wm,uc     &                      \\
3913.48   & 0.90 &  .47 Ti II-.294        &  .49 Re (3hw) wm,uc  \\
3915.40   & 0.90 &  .38 Re II (10h)       &                      \\
3926.27   & 0.95 &  .28 Re (4Hw) wm,uc    &                      \\
3939.34   & 0.95 &  .38 Re (5h) wm,uc     &                      \\
3947.45   & 0.90 &  .43 Re (10h) wm,uc    &                      \\
3952.11   & 0.98 &  .12 Fe II-.034        & .10 Re (3h) wm,uc    \\
4031.41cw & 0.88 &  .44 Fe II-.101        &   .42 Re II (20c6)   \\
4042.77   & 0.92 &  .78 Re (3) wm,uc      &                      \\
4043.09   & 0.97 &  .10 Re (3) wm,uc      &                      \\
4056.90   & 0.91 &  .91 N  I-.010         &   .91 Re (10h) wm,uc \\
4062.51br & 0.91 &  .40 Re (4h) whm uc    &                      \\
4069.10br & 0.82 &  .12 Re (30c.26Wl)wm,uc&                      \\
4091.96   & 0.91 &  .96 Re II (20h)       &   .94 Cr II-.005     \\
4135.41   & 0.95 &  .43 Re (2h) wm,uc     &                      \\
4146.77   & 0.97 &  .70 Re (5c.21Wl) wm,uc&    +                 \\
4149.51   & 0.96 &  .46 Re (5Hw) wm,uc    &                      \\
4231.51   & 0.95 &  .53 Re (20c.28Wl)wm,uc&                      \\
4240.22   & 0.96 &  .18 Re (4cw) wm,uc    &                      \\
4248.31   & 0.97 &  .27 Re (4h) wm,uc     &                      \\
4269.98   & 0.96 &  .94 Re (20c.39Wl)wm,uc&                      \\
4289.11   & 0.98 &  .09 Re (4h) wm,uc     &                      \\
4311.68   & 0.91 &  .68 Re II (20h)       &   .67 Mo II-.018     \\
4316.53   & 0.96 &  .52 Re (4h) wm,uc     &                      \\
4330.68   & 0.98 &  .79 Ti II-.100        &   .69 Re (9) wm,uc   \\
4356.29   & 0.99 &  .29 Re (6) wm,uc      &   .29 N I-.094       \\
4380.97   & 0.99 &  .00 Re II (10cwl)     &                      \\
4389.55   & 0.98 &  .60 Re II (10cwl)     &                      \\
4423.00   & 0.91 &  .02 Re II (20c)       &                      \\
4452.71db & 0.93 &  .68 Re II (10)        &   .82 Dy III-.005    \\
4473.31   & 0.93 &  .31 Re (20) wm,uc     &                      \\
4481.29   & 0.46 &  .32 Mg II-.315        &   .32 Re II (100c)   \\
4489.97   & 0.98 &  .99 Re (8c) wm,uc     &                      \\
4519.05   & 0.97 &  .09 Re (5) wm,uc      &                      \\
4520.96   & 0.95 &  .97 Re II (20c)       &                      \\
4584.50   & 0.98 &  .49 Re (30c.51Ws)wm,uc&    (remeasure v br)  \\
4596.54   & 0.91 &  .56 Re (100h) wm,uc   &                      \\
4673.26   & 0.87 &  .26 Si II-.205        &   .15 Re (100c) wm,uc\\
4690.92   & 0.98 &  .90 Re (10h) wm,uc    &                      \\
4703.61   & 0.98 &  .70 Re (10cw) wm,uc   &                      \\
4714.75   & 0.98 &  .78 Re (10) wm,uc     &                      \\
4724.65   & 0.98 &  .62 Re (4) wm,uc      &                      \\
4765.77   & 0.98 &  .81 Re (8h) wm,uc     &                      \\
4904.35   & 0.94 &  .33 Re II (10cw)      &                      \\
4909.74   & 0.97 &  .71 Re (4cw) wm,uc    &                      \\
5070.62   & 0.95 &  .58 Fe II-.161        &   .64 Re (3h) wm,uc  \\
5099.20   & 0.97 &  .20 Re (5c) wm,uc     &                      \\
5286.65   & 0.96 &  .68 Re (5h) wm,uc     &                      \\
\br
\end{tabular}
\end{center}
\end{table}
\normalsize

After tests with weak lines
in the original study of the rhenium spectrum by
Meggers (1952), it became clear that many unclassified rhenium
lines were also present in HD 65949.
Table~\ref{tab:re2} lists 59 stellar features that
are primarily or partially due to rhenium, mostly attributed
to \ion{Re}{ii}.
All of the unclassified laboratory lines
listed in our table are present in the spark spectrum, but
absent in the arc spectrum.

The first column of Table~\ref{tab:re2} gives the stellar
wavelength and the second the intensity at the line
core in units of the continuum.  The very weak feature
near $\lambda$3773 has Cent(i) of 1.00 due to a
slight misplacement of the continuum.  The columns
labeled ID$_1$ and ID$_2$ indicate primary (dominant)
and secondary contributors to the stellar feature.  Entries are
the laboratory
wavelength (fractional part of an angstrom only),
the spectrum, the intensity and description of the
laboratory line (e.g. $c$ means complex, $w$ wide, $db$ double, $br$
broad, etc.,
further details are in the reference cited.).
The ionization stage of lines identified as `Re'
is unclassified (uc). The feature at $\lambda$4146.77 is suspected
of having an unidentified contributor because of the
wavelength discrepancy. Intensity
parameters (IP) are given for non-\ion{Re}{ii} lines. Such lines
are estimated to be strong if IP is greater than 0.1.
Thus, \ion{Mo}{ii} $\lambda$3742.017\,\AA\ probably does not dominate
the nearby stellar feature, while the stellar feature at
4031.41\,\AA\, is mostly \ion{Fe}{ii}.

\subsection{Rhenium abundance}

Only one usable \ion{Re}{ii} line, $\lambda$3742.30, has a
modern oscillator strength (Palmeri, et al. 2005).
The measured equivalent width from our spectrum
is 74~\,m\AA. We have recalculated
the abundance, using
the revised model, but there remains an uncertainty due to
unknown hyperfine structure (hfs).  Laboratory spectra show
significant hfs for lines of \ion{Re}{i,\,ii}, due to the large nuclear
magnetic moment and electric quadrupole moment of rhenium. Measurements
of hfs for Multiplet UV1 of \ion{Re}{ii} have been presented by Wahlgren
et al. (1997). If we use a
microturbulence of 4 km s$^{-1}$ to attempt to allow
for the hfs, we find log(N$_{\rm Re}$/N$_{\rm total}$) = $-$6.45, which
corresponds to an excess of 5.36 dex over the solar
value. The hfs for optical lines will be measured in the near future
to investigate the potentially extraordinary abundance for
rhenium implied by our calculations.

\section{Osmium}

\begin{table}[h]
\caption{\label{tab:os2}Stellar and laboratory \ion{Os}{ii} lines}
\scriptsize
\begin{center}
\begin{tabular}{cccc}
\br
$\lambda^*(\AA )$&\rule{8mm}{0mm}${\rm ID_1}$&$\rm ID_2$\\
\mr
3692.69&.64 Os{\sc ii} (6)&          \\
3706.63&.64 Os{\sc ii} (1)&.55 Au II \\
3810.93&.92 Os{\sc ii} (1)&.90 P II  \\
3817.84&.84 Os{\sc ii} (1)&          \\
3848.78&.81 Os{\sc ii} (1)&          \\
3894.87&.89 Os{\sc ii} (2)&          \\
4050.13&.14 Os{\sc ii} (1)&.08 S II  \\
4109.08&.09 Os{\sc ii} (1)&          \\
\br
\end{tabular}
\end{center}
\end{table}

We claim a robust identification of \ion{Os}{ii}.  Stellar
and laboratory wavelengths are listed in Table~\ref{tab:os2}.
The format is similar to that of Table~\ref{tab:re2}.
The laboratory wavelengths are from van Kleef (1960), and
comprise all lines with intensities greater than unity
with the exception of $\lambda$4371.15, which was $not$
present at a detectible level in our spectra.

A wavelength coincidence statistics (WCS) run
(cf. Cowley \& Hensberge 1981) shows that finding eight out of
nine lines within 60 m\AA\, has a probability of less than
one in 5000 of occurring by chance.

{\it Unfortunately,
none of these \ion{Os}{ii} lines have oscillator strengths
in VALD}.  Thus we have no means to demonstrate that
$\lambda$4371.15 is below the threshold of detectibility,
though it is highly improbable that this is not the case.

\begin{figure}[t]
\includegraphics[width=4.0in,angle=-90]{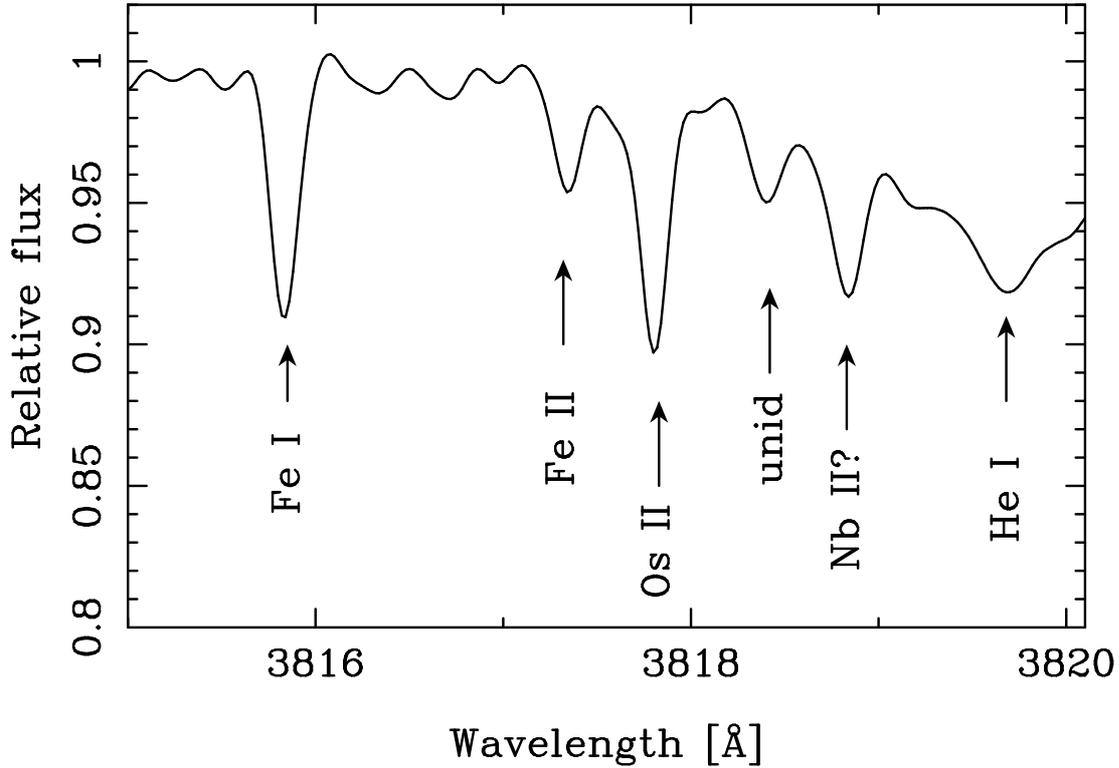}
\caption{\label{fig:re2}
Os {\sc ii} $\lambda$3817.84.  See url given in text for complete
identifications.}
\end{figure}

\section{Tellurium}

\ion{Te}{ii} is almost certainly present, though we cannot call
the identification ``robust.''  WCS places the coincidences
significant at the $<$ 0.001 level, but only the strongest
lines of Sansonetti \& Martin (2003) are clearly present.
A listing of observed lines
is given in Table~\ref{tab:te2}.  Intensities
and laboratory wavelengths (fraction of
an angstrom only) are from
Sansonetti \& Martin; {\it no transition probabilities are
given.}

\begin{table}[h]
\caption{\label{tab:te2}Stellar and laboratory \ion{Te}{ii} lines}
\scriptsize
\begin{center}
\begin{tabular}{lll}
\br
$\lambda^*(\AA )$& Cent(i)&\rule{8mm}{0mm}ID \\
\mr
4654.35&  0.97&   .37 Te II (900)  \\
5576.40&  0.95&   .35 Te II (800)    \\
5649.25&  0.95&   .26 Te II (800)     \\
5666.20&  0.97&   .20 Te II (500)    \\
5708.11&  0.95&   .12 Te II (1000)   \\
5755.86&  0.97&   .85 Te II (800)    \\
5974.70&  0.98&   .70 Te II (500)    \\
\br
\end{tabular}
\end{center}
\end{table}

\section{Mercury}

Mercury abundances were recalculated for relatively weak
\ion{Hg}{i,ii} lines using a cooler
model atmosphere than was used in Paper I. These lines
contain relatively little isotopic information. We noted previously that
the strong $\lambda$3984 line indicates that the heavy
isotope $^{204}$Hg must be a major contributor to the feature.

Two \ion{Hg}{i} lines ($\lambda\lambda$4358 and 5460)
give log(N$_{\rm Hg}$/N$_{\rm total}$) = $-$4.94, while \ion{Hg}{ii}
($\lambda\lambda$6149 and 7944) gives $-$4.46.
The first and second spectra now disagree by
0.48 dex, while in Paper I, the difference was
only 0.22 dex.  Abundances from the neutral and
ion states might be reconciled if an intermediate temperature
were adopted. A stratified mercury abundance can also reconcile
the differences.

\section{Additional rare spectra}

Paper I discussed the unambiguous identification of \ion{Pt}{ii},
and pointed out that the heavy platinum isotope $^{198}$Pt
probably dominated.  \ion{Xe}{ii} is also clearly present.  Less
definite, though probably present are \ion{Kr}{ii}, \ion{Nb}{ii},
\ion{Ru}{ii}, \ion{Sb}{ii}, \ion{Pr}{iii}, \ion{Ho}{iii} and \ion{Au}{ii}.
No second spectra of the lanthanides are identified, probably due to the
high temperatures in the stellar atmosphere.

\section{The mysterious, unidentified feature at $\lambda$3859}

Since Adelman's (1973) extensive study, numerous uranium
abundances in magnetic Ap stars
have been claimed to be determined using a feature
near the position of \ion{U}{ii} $\lambda$3859.5716 (Sansonetti
\& Martin 2003).  However, the stellar position has
been generally measured at a slightly longer wavelength.
In the spectrum of magnetic Ap stars, the feature is often complex,
and could accommodate both \ion{U}{ii} as well as a blending
feature.  Cowley \& Arnold (1978) discussed several
possibilities for the blends.
We measured an equivalent width of 24~\,m\AA\ for the line at
3859.63\,\AA.  The line
is symmetrical, and appears on all four FEROS exposures.
It is unlikely to be any of the features suggested by
Cowley \& Arnold because of the temperature and
abundances of HD 65949.  We still cannot
identify it, but suggest it could be involved in the
blend seen in the cooler magnetic Ap stars.  There is
considerable interest in this feature, since it is
used in cosmochronology (cf. Cayrel, et al. 2001).

\section*{References}
\begin{thereferences}
\item Abt H A and Morgan W W 1969 {\it AJ} {\bf 74} 813
\item Adelman S J 1973 {\it ApJS} {\bf 26} 1
\item Castelli F and Hubrig S 2007 {\it A\&A} {\bf 475} 1041
\item Cayrel R et al. 2001 {\it Nature} {\bf 409} 691
\item Cowley C R and Arnold C E 1978 {\it ApJ} {\bf 226} 420
\item Cowley C R and Hensberge H 1981 {\it ApJ} {\bf 244} 252
\item Cowley C R, Hubrig S, Gonz\'{a}lez, G. F. and Nu\~{n}ez, N.
2006, {\it A\&A} {\bf 455} L21 (Paper I)
\item Fuhr J R and Wiese W L 2006 {\it J. Phys. Chem. Ref. Data}
{\bf 35} 669
\item Hubrig S, North P, Sch\"{o}ller M and Mathys G
2006 {\it AN} {\bf 327} 289
\item Kupka F G, Ryabchikova T A, Piskunov N E, Stempels
H C and Weiss W W 2000, {\it Baltic Astron.} {\bf 9} 590
\item Meggers W F 1952 {\it J. Res. NBS} {\bf 49} 187
\item Meggers W F, Catal\'{a}n M A and Sales M 1958 {\it J. Res.
NBS} {\bf 61} 441
\item Muthsam H and Cowley C R 1984 {\it A\&A} {\bf 130}
348
\item Palmeri P, Quinet P, Bi\'{e}mont \'{E}, Xu H L
and Svanberg S 2005 {\it MNRAS} {\bf 362} 1348
\item Sansonetti J E and Martin W C 2003 {\it Handbook of
Basic Atomic Spectroscopic Data}
\verb=http://physics.nist.gov/PhysRefData/Handbook/index.html=
\item van Kleef Th A M 1960 {\it Proc. Koninkl. Ned. Akad.
Wetenschap B.} {\bf 63} 501
\item Wahlgren G M, Johansson S G, Litz\'en U et al. 1997 {\it ApJ} {\bf
475} 380
\end{thereferences}
\end{document}